\documentclass[12pt]{article}

\usepackage{scicite}
\usepackage{times}
\usepackage{amssymb}
\usepackage{graphicx}
\usepackage{dcolumn}
\usepackage{bm}
\usepackage{amsmath}
\usepackage{mathrsfs}
\usepackage{textcomp}
\usepackage{upgreek}

\def\be{\begin{equation}}
\def\ee{\end{equation}}
\def\bea{\begin{eqnarray}}
\def\eea{\end{eqnarray}}

\newenvironment{sciabstract}{%
\begin{quote} \bf}
{\end{quote}}

{\title{Observation and quantification of pseudogap in unitary Fermi gases}}

\author
{{Xi Li}$^{1,2\dagger}$, Shuai Wang$^{1,2\dagger}$, Xiang Luo$^{1,2}$, Yu-Yang Zhou$^{1,2}$,
\\
Ke Xie$^{1,2}$, Hong-Chi Shen$^{1,2}$, Yu-Zhao Nie$^{1,2}$, Qijin Chen$^{1,2,3}$,
\\
Hui Hu$^{4,1}$,Yu-Ao Chen$^{1,2,3}$, Xing-Can Yao$^{1,2,3}$ \& Jian-Wei Pan$^{1,2,3}$\\
\\
\normalsize{$^{1}$ Hefei National Research Center for Physical Sciences at the Microscale}\\
\normalsize{and School of Physical Sciences,}\\
\normalsize{University of Science and Technology of China, Hefei 230026, China}\\
\normalsize{$^{2}$Shanghai Research Center for Quantum Science}\\
\normalsize{and CAS Center for Excellence in}\\
\normalsize {Quantum Information and Quantum Physics,}\\
\normalsize{University of Science and Technology of China, Shanghai 201315, China}\\
\normalsize{$^{3}$Hefei National Laboratory,}\\
\normalsize{University of Science and Technology of China, Hefei 230088, China}\\
\normalsize{$^{4}$Centre for Quantum Technology Theory,}\\
\normalsize {Swinburne University of Technology, Melbourne 3122, Australia}\\
}

\date{}

\begin{document}

\baselineskip24pt

\maketitle

\begin{sciabstract}

   The nature of pseudogap lies at the heart of strongly-interacting superconductivity and superfluidity. 
   With known pairing interactions, unitary Fermi gases provide an ideal testbed to verify whether a pseudogap can arise from many-body pairing. 
   Here we report the observation of the long-sought pair-fluctuation-driven pseudogap in homogeneous unitary Fermi gases of lithium-6 atoms, by precisely measuring the spectral function through momentum-resolved microwave spectroscopy without the serious effects of final-state effect. 
   We find a large pseudogap above the superfluid transition. 
   The inverse pair lifetime exhibits a thermally-activated exponential behavior, uncovering the microscopic virtual pair breaking and recombination mechanism. The obtained large, $\bm{T}$-independent single-particle scattering rate is comparable with that set by the Planckian limit. Our findings quantitatively characterize the pseudogap in strongly-interacting Fermi gases, highlighting the role of preformed pairing as a precursor to superfluidity.

\end{sciabstract}

\section*{Introduction}

    The understanding of the pseudogap phase, a very abnormal ``normal'' state of cuprate superconductors, is widely believed to be the key to unlocking the mysterious microscopic mechanism of high-$T_\text{c}$ superconductivity\cite{Ding1996,Loeser1996}. 
    Intuitively, the pseudogap, manifested as a gap in the normal-state single-particle energy spectrum, would arise from strong pair fluctuations, as a precursor of coherent pair condensation beyond the standard Bardeen-Cooper-Schrieffer (BCS) theory\cite{Chen2005}. 
    However, in cuprate superconductors, the confirmation of such a simple scenario is hindered by some competing quantum orders, such as the d-density wave\cite{Chakravarty2001}, the pair-density wave\cite{Fradkin2015} and the stripe phase\cite{Kivelson2003}, particularly in the underdoped region\cite{Keimer2015,Damascelli2003}.

    Unitary Fermi gases, featuring a diverging $s$-wave scattering length\cite{Chin2010}, provide an ideal platform and quantum simulator\cite{Bloch2012} for observing the long-sought pair-fluctuation-driven pseudogap\cite{Stajic2004}, due to their unprecedented controllability and purity\cite{Giorgini2008,Zwerger2012,Randeria2014}. 
    To this end, radio-frequency (rf) spectroscopy\cite{Chin2004,Schunck2008,Murthy2018} was developed to probe the pairing of fermions; particularly, with momentum resolution, the single-particle spectral function\cite{Stewart2008,Gaebler2010,Feld2011} was measured. 
    Unfortunately, the experimental resolution was insufficient to resolve the two expected quasiparticle branches. 
    Instead, only a very broad spectral response was observed, with the dispersion exhibiting a back-bending behavior at wavenumber $k$ larger than the Fermi wavenumber $k_\text{F}$\cite{Stewart2008,Gaebler2010}. 
    This back-bending behavior was attributed to a BCS-like quasiparticle dispersion, providing primitive indication of a pseudogap\cite{Gaebler2010}. 
    However, there has been an on-going debate regarding this interpretation\cite{Zwerger2012,Randeria2014,Mueller2017}, since the back-bending at $k>k_\text{F}$ could occur in any short-range interacting Fermi systems\cite{Schneider2010} and the observed spectra have also been argued to be interpretable by a Fermi-liquid description\cite{Nascimbene2011,Mueller2017,Zwerger2012}. 
    
    In this work, we prepare a homogeneous unitary Fermi gas of $^{6}$Li atoms in a cylindrical box trap\cite{Gaunt2013,Mukherjee2017,Baird2019,Li2022}, and develop a novel momentum-resolved microwave spectroscopy to probe the fermion spectral function $A(\bm{k},\omega)$. 
    On the one hand, the box trap eliminates the trap inhomogeneity which prohibited the extraction of a homogeneous spectral function in previous experiments. 
    On the other hand, the microwave spectroscopy removes the complications associated with the final-state interactions\cite{Baym2007,Schunck2008} in the widely adopted rf spectroscopy\cite{Schunck2008,Mukherjee2019}, particularly for $^{6}$Li atoms, such as spectral shift and broadening. 
    Here the final atoms in the high-lying hyperfine level are essentially free from interactions with the initial atoms in the many-body system. 
    Although the microwave transition is sensitive to magnetic field variations, we can achieve an ultrahigh stability of the magnetic field and thus realize this unique microwave transition with high energy-resolution 

    We simultaneously observe the two BCS-like quasiparticle branches in the spectral function, both below and above the superfluid transition temperature $T_\text{c}$, which is a direct manifestation of pairing. 
    The obtained large $\Delta/E_\text{F}$ above $T_\text{c}$ unambiguously reveals the existence of a pair-fluctuation-driven pseudogap, where $\Delta$ is the pairing (pseudo)gap and $E_\text{F}$ is the Fermi energy. 
    The pseudogap persists up to the highest temperature ($1.51T_\text{c}$) that we have achieved, indicating a sizable pseudogap window in the unitary Fermi gas. 
    Furthermore, by analyzing the energy distribution curve (EDC) in $A(\bm{k},\omega)$, a thermally activated exponential $T$ dependence of the inverse pair lifetime $\Gamma_0$ is revealed, with an activation energy of $2\Delta_0$, where $\Delta_0$ is the low-$T$ pairing gap. 
    This underlies the mechanism that the pair lifetime is dominated by a virtual pair breaking and recombination process. 
    The single-particle scattering rate $\Gamma_1$ reaches the universal Planckian limit\cite{Zaanen2004,Grissonnanche2021,Bruin2013} of $\sim k_\text{B} T/\hbar$ near $T_\text{c}$, where $k_\text{B}$ and $\hbar$ are the Boltzmann and reduced Planck constants, respectively. 
    Our result settles the long-standing debates on the existence of a pseudogap in unitary Fermi gases, which provides crucial information in establishing a proper theory of strongly-interacting Fermi superfluid, and supports pairing as a possible origin of the pseudogap in high-Tc superconductors, within the framework of preformed-pair superconductivity.

    \section*{Experimental scheme and setup}

    \begin{figure*}[h!]
        \centering
        \includegraphics[width =0.8\linewidth]{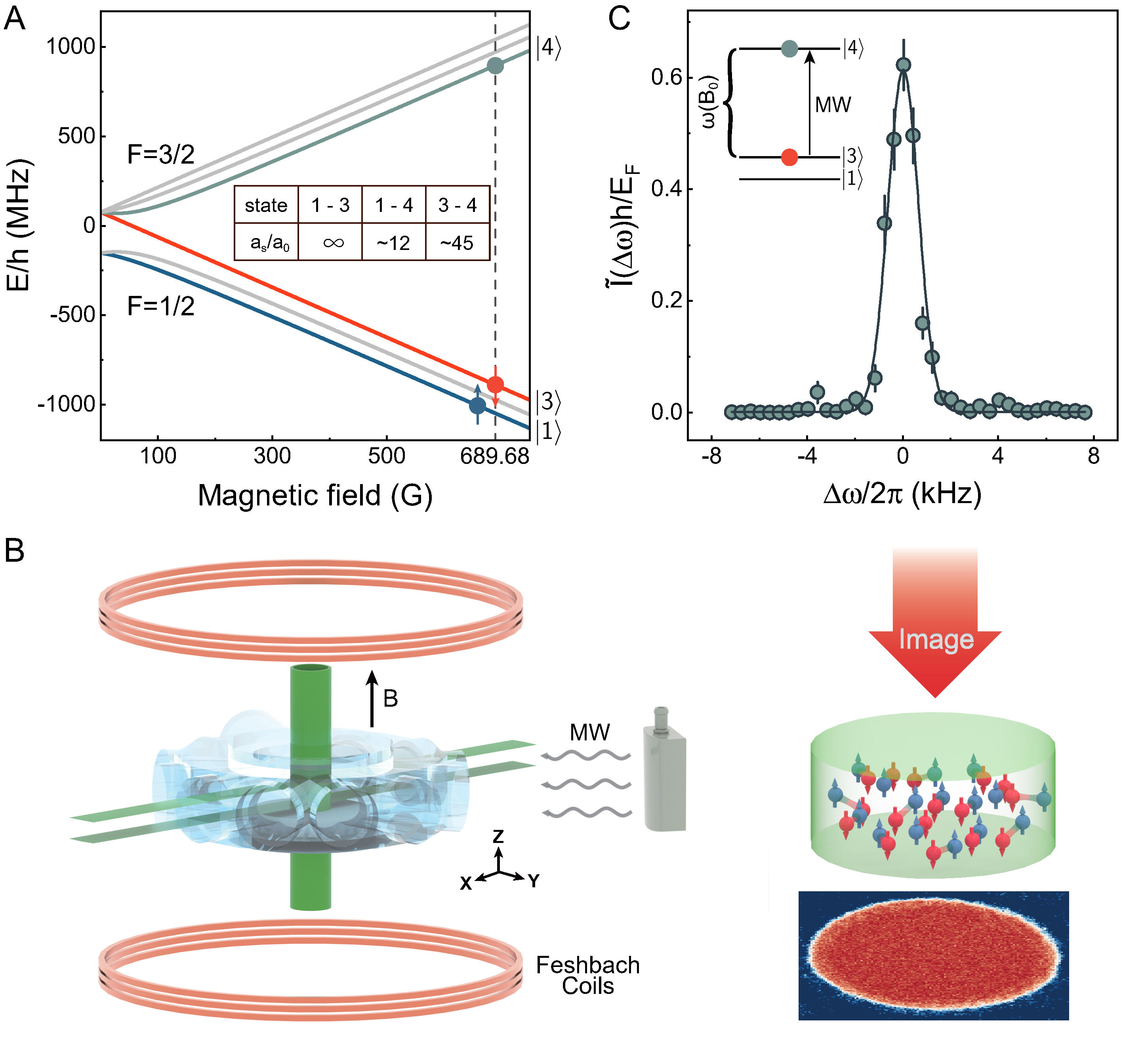}
        \caption{\textbf{Experimental scheme}.
        (\textbf{A}) Magnetic field dependence of the $2^{2}S_{1/2}$ ground state of $^6$Li atoms. 
        The dashed line marks the applied $|3\rangle\rightarrow|4\rangle$ microwave transition at the Feshbach resonance $B_0=689.68$~G. 
        The inset table shows the $s$-wave scattering lengths for $|1\rangle-|3\rangle$, $|1\rangle-|4\rangle$, and $|3\rangle-|4\rangle$ collisions. 
        (\textbf{B})	Sketch of the experimental setup. 
        Left panel: The cylindrical box trap consists of a tube and two sheets of 532~nm laser beams, which are generated by two spatial light modulators. 
        The magnetic field is generated by a pair of Feshbach coils. 
        The microwave pulse traveling along the horizontal plane is generated by an antenna. 
        Right panel: The incident imaging laser along the vertical direction gives an \textit{in situ} image of the unitary Fermi gas at $T_\text{c}$.
        (\textbf{C}) Microwave (MW) spectrum of an ideal Fermi gas. 
        The data point (dark green circle) corresponds to an average over approximately 30 independent measurements. 
        The error bar represents one standard deviation. 
        The dark green solid line is the fit to a Gaussian function. 
        The inset shows a sketch of the measurement.}
        \label{fig1}
    \end{figure*}
 
    The experimental procedure of creating a homogeneous unitary Fermi gas at a given temperature follows our previous work\cite{Li2022}, except for two main distinctions. 
    First, $^{6}$Li atoms are equally prepared in the two hyperfine levels $|1\rangle \equiv | F = 1/2, m_F = 1/2 \rangle$ and $|3\rangle \equiv | F=3/2, m_F = -3/2 \rangle$ at the magnetic field $B_0=689.68$~G, where inter-spin interactions are resonant and unitarily limited\cite{Zuern2013}. 
    Second, a cylindrical box trap with a diameter of 84~$\upmu$m and a height of 41~$\upmu$m is formed by a hollow cylinder and two sheets of repulsive 532~nm laser beams, as depicted in Fig.~\ref{fig1}B. 
    After the initial state preparation, approximately $2.7\times10^{6}$ atoms at $0.26T_\text{F}$ are confined in the cylindrical box trap\cite{supple}, where $T_\text{F}=E_\text{F}/k_\text{B}$ is the Fermi temperature. 
    We then adiabatically lower the potential depth in 0.6~s and hold the box trap for an additional 0.5~s to reach thermal equilibrium. 
    For a homogeneous unitary Fermi gas at $T_\text{c} \simeq 0.17T_\text{F}$ (e.g., see the in-situ image in Fig.~\ref{fig1}B), the realized density is $n \simeq 1.09\times 10^{13}$~$\text{cm}^{-3}$, yielding a large $E_\text{F}= \hbar^2k_\text{F}^2/2m\simeq 2\pi\hbar\times 39.5$~kHz. Here, $m$ is the atomic mass and $k_\text{F} = (3 \pi^2 n)^{1/3}$. 
    To determine the temperature $T$ at a given final depth, we measure the energy $E$ through Bragg spectroscopy in the long-wavelength limit\cite{Li2022} and compare it with the known equation of state (EoS)\cite{supple,Ku2012}.
    As an independent validation of our EoS thermometry, the pair momentum distribution is also measured through the combination of an interaction quench to the molecular side of the Feshbach resonance and a matter-wave focusing technique\cite{Li2022,supple}, which unambiguously reveals the onset of pair condensation at $T_\text{c}$ (see Fig.~S1). 
   
    In our momentum-resolved microwave spectroscopy, atoms in level $|3\rangle$ are transferred to an initially unoccupied hyperfine level $|4\rangle \equiv | F=3/2, m_F = -1/2 \rangle$ by using a Gaussian-shaped microwave pulse with an intensity of $V_\text{I}(t)=V_0\mathrm{e}^{-t^2/2\tau^2}$ at the time interval $-t_\mathrm{p}/2 \le t \le +t_\mathrm{p}/2$.  
    Here, $V_0=\hbar\Omega_\mathrm{m}/2$ is the maximal coupling strength of the pulse, characterized by a peak Rabi frequency $\Omega_\text{m}$ of approximately $2\pi\times 1.35$~kHz, $t_\mathrm{p}=850~\upmu$s is the pulse duration, and $\tau=180~\upmu$s represents the $1/\sqrt{e}$ half width of the pulse. 
    We then immediately turn off the box trap, let the gas expand ballistically in the residual magnetic curvature for 5~ms, and finally measure the density distribution $n_\text{2D}(x,y)$ of atoms in level $|4\rangle$ using absorption imaging with a high signal-to-noise ratio (SNR) and a high spatial resolution\cite{supple}.
    We determine the transferred atom number $N_4 (\Delta\omega)=\int n_\text{2D}(x,y)\,\text{d}x\text{d}y$ as a function of the microwave frequency offset $\Delta\omega=\omega_\text{MW}-\omega(B_0)$, where $\omega_\text{MW}$ is the absolute microwave frequency and $\omega(B_0)\simeq 2\pi\times 1787.398$~MHz corresponds to the Zeeman energy splitting $E_{3-4}(B_0)=\hbar\omega(B_0)$ between the $|3\rangle$ and $|4\rangle$ hyperfine levels at $B_0=689.68$~G.
    According to Fermi's golden rule, within the linear response, $N_4(\Delta\omega)$ is proportional to $N_3\Omega_\text{m}^2\tau$, where $N_3$ is the initial atom number in level $|3\rangle$\cite{supple}. 
    Thus, we define a dimensionless microwave spectrum $\tilde{I}(\Delta\omega) = [N_4(\Delta\omega)/N_3](2E_\text{F}/\pi^{3/2}\hbar\Omega_\text{m}^2\tau)$, which is normalized in accordance with $\int \tilde{I}(\Delta\omega)\,\text{d}(\hbar \Delta\omega/E_\text{F})=1$.   
    
    The use of the hyperfine level $|4\rangle$ as the final state brings two key advantages due to the negligible $s$-wave scattering lengths $a_\text{14}$ and $a_\text{34}$ (see the inset table of Fig.~\ref{fig1}A): 
    (i) the final-state interaction can be neglected, i.e., $k_\text{F}a_\text{14}\simeq0.0043\simeq0$; 
    and (ii) the isotropic momentum information $n(\bm{k}, \Delta\omega)$ carried by the outcoupled atoms in level $|4\rangle$ can be accurately extracted through ballistic expansion. 
    More specifically, we perform an inverse Abel transform\cite{Stewart2008} on the measured $n_\text{2D}(x,y)$ to reconstruct the full density distribution $n_\text{3D}(r)$ (see Fig.~S4) and then obtain the momentum distribution via appropriate variable substitution\cite{supple}, owing to the scaling $r \propto k$ satisfied in the ballistic expansion.
    Due to the negligible momentum of the microwave photon, we have $n(\bm{k},\Delta\omega)\propto A(\bm{k},\omega)f(\omega)$\cite{supple,Chen2009a}, where the single-particle excitation energy $\omega = \epsilon_{\bm{k}} / \hbar-\Delta\omega$ is offset by the free-particle kinetic energy $\epsilon_{\bm{k}} = \hbar^2 \bm{k}^2/2m$, as a result of energy conservation. 
    Thus, the single-particle spectral function $A(\bm{k},\omega)$ can be determined by performing proper transformations on $n(\bm{k},\Delta\omega)$ and dividing by the Fermi function $f(\omega)=1/[\mathrm{e}^{(\hbar \omega-\mu)/k_\text{B}T}+1]$\cite{supple}, where the chemical potential $\mu$ can be accurately obtained from the known thermodynamic EoS\cite{Li2022, Ku2012} for the given reduced temperature $T/T_\text{F}$. 
    
    Unlike the radio-frequency transition $|3\rangle\to|2\rangle$ used in previous experiments\cite{Murthy2018,Mukherjee2019}, the realization of a high energy-resolution microwave transition $|3\rangle\to|4\rangle$ represents a great experimental challenge. 
    This is because at $B_0=689.68$~G, the energy gradient $\delta E_{3-4}(B)/\delta B$ is approximately $2\pi\hbar\times 2.79$~MHz/G. 
    In comparison with $\delta E_{3-2}(B)/\delta B\simeq 2\pi\hbar\times $8.48~kHz/G, we need to significantly improve the magnetic field stability by at least two orders in magnitude, to the level of 0.1 ppm (parts per million).   
    To this end, we develop and combine several methods to stabilize the magnetic field, e.g., passive magnetic field shielding, ultrastable current sources, active magnetic field compensation, etc (see Fig.~S2). 
    To evaluate the final stability of the magnetic field, two complementary measurements are implemented for an ideal Fermi gas, where all the atoms are initially prepared in level $|3\rangle$. 
    On the one hand, Rabi oscillations between levels $|3\rangle$ and $|4\rangle$ are measured (see Fig.~S3), from which an unprecedented small uncertainty $\delta B$ of approximately 20~$\upmu$G at $B_0=689.68$~G is inferred\cite{supple}. 
    On the other hand, as shown in Fig.~\ref{fig1}C, the microwave spectrum $\tilde{I}(\Delta\omega)$ is Fourier-limited.
    We obtain a very high energy-resolution of $1.51(5)$~kHz for the microwave transition (i.e., approximately $0.038 E_\text{F}$), by fitting the data to a Gaussian function. 
    
\section*{Single-particle spectral function}

\begin{figure*}[h!]
        \centering
        \includegraphics[width =0.9\linewidth]{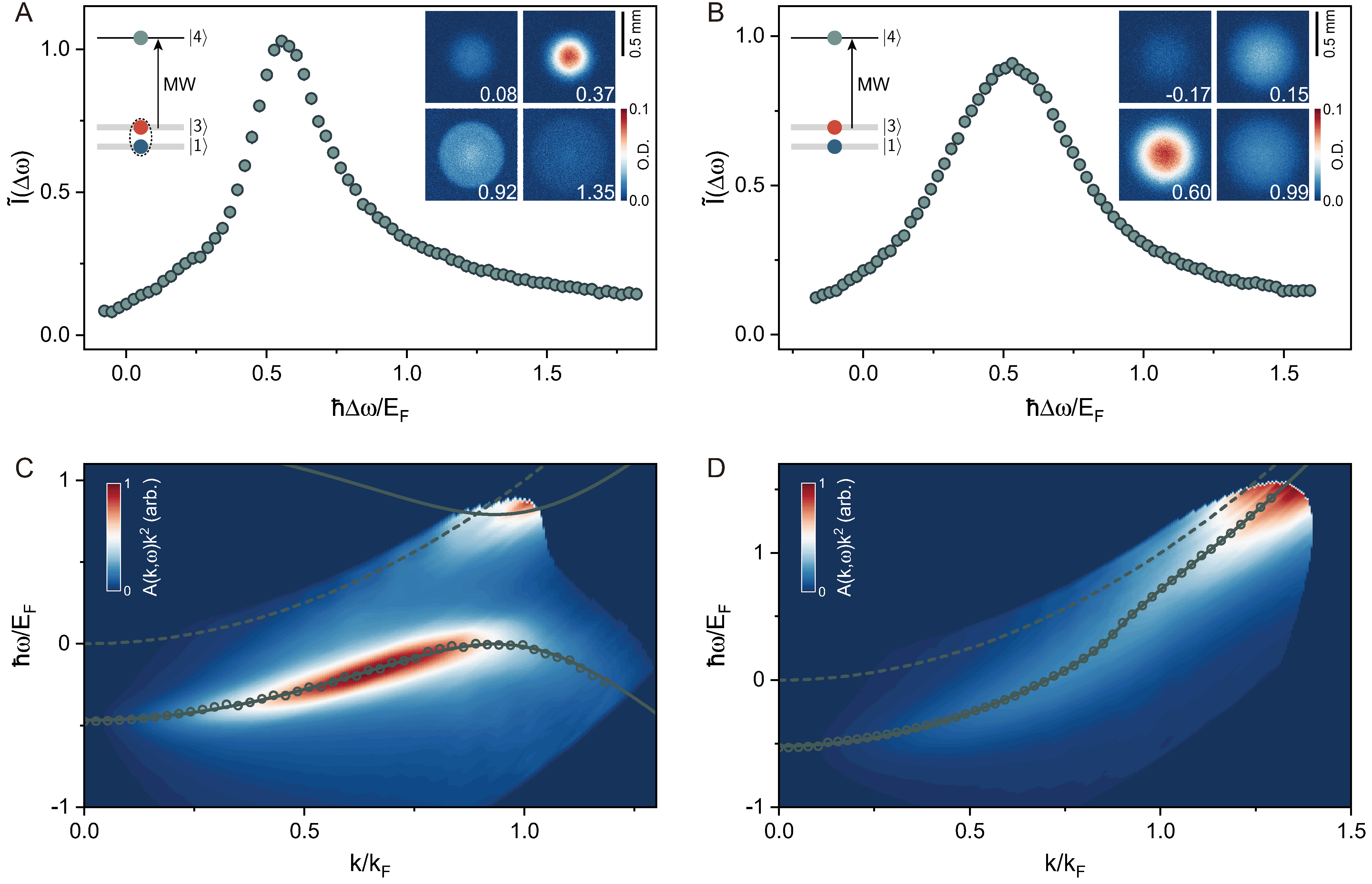}
        \caption{\textbf{Microwave spectra at $0.77\boldmath{T}_\text{c}$ and $1.51\boldmath{T}_\text{c}$}.
        (\textbf{A}) and (\textbf{B}) The dimensionless microwave rate $\tilde{I}(\Delta\omega)$, defined in the text, as a function of the normalized frequency offset $\hbar\Delta\omega/E_\text{F}$ at $T=0.77T_\text{c}$ and $1.51T_\text{c}$, respectively. 
        Each data point (blue circle) represents an average of approximately 100 independent measurements with the standard statistical error being calculated. 
        In each plot, the left inset shows the experimental scheme, where the broadened energy levels (continuous spectra) of $|1\rangle$ and $|3\rangle$ are due to strong interatomic interactions. 
        The dashed ellipse denotes condensed pairs below $T_\text{c}$. 
        The right inset exhibits the measured $n_{\mathrm{2D}}(x,y)$ at four exemplary normalized frequency offsets, to highlight a pronounced $\Delta\omega$ dependence. 
        (\textbf{C}) and (\textbf{D}) Intensity plots of the momentum-resolved microwave spectra $A(\bm{k},\omega)\bm{k}^2$ at $T=0.77T_\text{c}$ and $1.51T_\text{c}$, respectively, as a function of the momentum $k=|\bm{k}|$ (normalized to $k_\text{F}$) and the single-particle excitation energy $\omega$ (normalized to $E_\text{F}$). 
        Note that, we do not show the data point with signal intensity less than approximately 0.005 times the maximum intensity, which is essentially the background noise and sets the boundary of the intensity plot. 
        Here, the circles are the maximal spectral response $E_{\mathrm{max}}(k)$. 
        At $T=0.77T_\text{c}$ in (\textbf{C}), where the two branches are clearly distinguishable, we show $E_{\mathrm{max}}(k)$ of the lower branch only. 
        The gray solid lines are fitting curves to the loci of $E_{\mathrm{max}}(k)$ using $E_\text{k}^{(\pm)}$ in (\textbf{C}) and $E_\text{k}$ in (\textbf{D}). 
        For comparison, the free-particle dispersion relation $\epsilon_{\bm{k}}$ is shown as the gray dashed line.}
        \label{fig2}
    \end{figure*}

    The ability to create a homogeneous unitary Fermi gas with a large Fermi energy and the realization of momentum-resolved microwave spectroscopy with a negligible final-state effect enable us to measure the single-particle spectral function $A(\bm{k},\omega)$ in previously inaccessible parameter regimes. 
    We first consider two limiting cases, i.e., unitary Fermi gases prepared at the lowest ($0.77T_\text{c}$) and the highest temperatures ($1.51T_\text{c}$). 
    The advantage of eliminating the final-state effect can be seen from the normalized dimensionless microwave rate $\tilde{I}(\Delta\omega)$ at these two temperatures, as reported in Fig.~\ref{fig2}, A and B.     

    In comparison with a similar recent measurement\cite{Mukherjee2019} but using the $|3\rangle\to|2\rangle$ transition, two distinctive features are observed in our microwave spectra: (i) The spectral width $E_\text{w}$, i.e., the full width at half maximum of the spectrum, is narrower than previous results, which indicates a larger spectroscopic pair size $\xi$, according to the relation $\xi^2\sim\hbar^2/mE_\text{w}$\cite{Schunck2008}. 
    Furthermore, our measured $E_\text{w}$ increases by only $\sim 0.2 E_\text{F}$ as temperature $T$ increases from $0.77T_\text{c}$ to $1.51T_\text{c}$, implying that the pair size of fermions does not change much over a wide range of temperatures, especially at $T>T_\text{c}$. 
    (ii) The peak position $E_\text{p}$ of the spectrum, which is mainly determined by $\Delta$ and the Hartree energy $U$\cite{Haussmann2009}, remains almost unchanged, i.e., $E_\text{p}\approx0.56E_\text{F}$ ($0.53E_\text{F}$) at $0.77T_\text{c}$ ($1.51T_\text{c}$). 
    This suggests that, as $T$ increases, $U$ may increase to compensate for the decrease in $\Delta$. 
    The weak temperature dependence of both spectral width $E_\text{w}$ and peak position $E_\text{p}$ implies a smooth change in the many-body pairing gap $\Delta$ across $T_\text{c}$. 
    We note that, the spectral response falls off as $\tilde{I}(\Delta\omega\rightarrow \infty)\propto C/(\Delta\omega)^{3/2}$ at $\hbar\omega\gg E_\text{F}$, where $C$ is the Tan's contact\cite{Schneider2010,Carcy2019,Mukherjee2019} characterizing short-range pair correlations. 
    However, a quantitative verification of such a $(\Delta\omega)^{-3/2}$ decay is difficult here, since a high SNR measurement can be achieved only in the range of $\hbar\Delta\omega<2 E_\text{F}$. 
    This is because, unlike previous experiments in which the pulse strength increased at high frequencies\cite{Stewart2008,Mukherjee2019}, we apply the same microwave intensity $V_\text{I}(t)$ for all the frequencies to avoid complications in analyzing the spectra. 
    
    To directly extract the many-body pairing gap, we present the momentum-resolved microwave spectra $A(\bm{k},\omega)\bm{k}^2$ at $T=0.77T_\text{c}$ and $1.51T_\text{c}$ in Fig.~\ref{fig2}, C and D, respectively. 
    For $T=0.77T_\text{c}$, we observe two well-separated excitation branches, despite the relatively weak signal of the upper branch, which arises from thermally excited quasiparticles\cite{Chen2009}. 
    In particular, the spectral weight near the Fermi surface (i.e., $k = |\bm{k}| \sim k_{\text{F}}$ and $\hbar\omega \sim \mu$) is significantly depleted, unambiguously revealing the existence of a pairing gap $\Delta$. 
    The observation of two quasiparticle branches and the direct visualization of the pairing gap highlight the technical advantage of our momentum-resolved microwave spectroscopy. 
    We emphasize that in all previous measurements of momentum-resolved rf spectroscopy\cite{Stewart2008,Gaebler2010,Feld2011}, only the lower quasiparticle branch was observed, and the data were plagued by the presence of trap inhomogeneity. 
    The identification of the pairing gap relies heavily on its back-bending behavior at an unphysically large wavenumber $k>k_\text{F}$. 
    This has caused serious doubts about the pairing interpretation of the back-bending behavior\cite{Nascimbene2011,Schneider2010}.
    
    With a much improved SNR in a homogeneous setting, we observe the back-bending at $k \simeq 0.93k_\text{F}$, which is lower than $k_\text{F}$ as expected. 
    Here, we quantitatively extract the pairing gap $\Delta$, by identifying the locus of the maximal spectral response $E_\text{max}(k)$ of the lower excitation branch (see circles in Fig.~\ref{fig2}C) with a BCS-like dispersion $E_\text{k}^{(-)}$\cite{supple,Haussmann2009}:
    
    \begin{equation}
        E_\text{k}^{(\pm)}=\mu\pm\sqrt{\left( \frac{\hbar^2k^2}{2m^*}+U-\mu \right)^2+\Delta^2}.
        \label{DispersionBCS}
    \end{equation}
    Here, $E_\text{k}^{(+)}$ and $E_\text{k}^{(-)}$ are the quasiparticle energy dispersions in the upper and lower branches, respectively, with $m^*$ being the effective mass. 
    The Hartree energy $U$ leads to a downshift of the spectrum from the free-particle dispersion (see the dashed line in Fig.~\ref{fig2}C). 
    The fitting curve to $E_\text{max}(k)$ with $m^*$, $U$ and $\Delta$ as fitting parameters is shown as the lower solid line in Fig.~\ref{fig2}C, which is in excellent agreement with the experimental data. 
    The obtained pairing gap $\Delta$ is $0.396(2)E_\text{F}$ at $0.77T_\text{c}$, slightly smaller than the most recent experimental result of $\Delta = 0.47(1)E_\text{F}$ from Bragg spectroscopy\cite{Biss2022}. 
    The fit also gives $m^*=1.118(6) m$ and $U=-0.374(3)E_\text{F}$. 
    Using these fitting parameters, $E_\text{k}^{(+)}$ is plotted as the upper solid line in Fig.~\ref{fig2}C, which reasonably describes the observed upper quasiparticle branch. 

    At sufficiently high temperatures, one expects a single excitation branch with a quadratic dispersion\cite{Chen2009,Haussmann2009} in $A(\bm{k},\omega)$. 
    We indeed observe a single upward branch for $T=1.51T_\text{c}$, as shown in Fig.~\ref{fig2}D. 
    However, the locus of $E_\text{max}(k)$ exhibits an ``S-shape'' dispersion, which is particularly evident near $k \sim k_\text{F}$. 
    This can be attributed to the existence of a rather small pairing gap at $1.51T_\text{c}$. 
    The two quasiparticle branches with BCS-like dispersions $E_\text{k}^{(\pm)}$ still exist; however, their thermal broadening smears out the spectral gap. 
    Nevertheless, as the wavenumber $k$ increases across the Fermi surface, the maximal spectral response evolves from the lower to the upper branch, thus resulting in an ``S-shape'' dispersion. 

    This picture motivates us to explain the ``S-shape'' dispersion with a phenomenological model: ${E_\text{k}} =\alpha_\text{k}^2E_\text{k}^{(+)}+\beta_\text{k}^2E_\text{k}^{(-)}$, where we approximate $\alpha_\text{k}^2,\beta_\text{k}^2=1/2 \pm \tanh[(k-k_\text{m})/\sigma]/2$. 
    Here, $k_\text{m} \sim k_\text{F}$ is the crossover point, and $\sigma$ controls the crossover rate between the two branches. 
    At small wavenumber $k \ll k_\text{F}$,  $\beta_\text{k}^2 \simeq 1$ and $\alpha_\text{k}^2 \simeq 0$, so that $E_\text{k}$ reduces to $E_\text{k}^{(-)}$. 
    Conversely, $E_\text{k} \simeq E_\text{k}^{(+)}$ holds at large wavenumber. 
    Moreover, at the crossover point, $\alpha_\text{k}^2=\beta_\text{k}^2=1/2$ indicates the equal contribution of the two branches to the maximal spectral response. 
    By using the phenomenological dispersion to fit the observed $E_\text{max}(k)$, as shown by the solid line in Fig.~\ref{fig2}D, we find a small pairing gap of $0.05(4)E_\text{F}$, as anticipated. 
    A larger Hartree shift of $U=-0.517(3)E_\text{F}$ is also extracted, which leads to an essentially unchanged peak position in $\tilde{I}(\Delta\omega)$ (see Fig.~\ref{fig2}, A and B), as we speculated above.

\begin{figure*}[h!]
    \centering
    \includegraphics[width =0.8\linewidth]{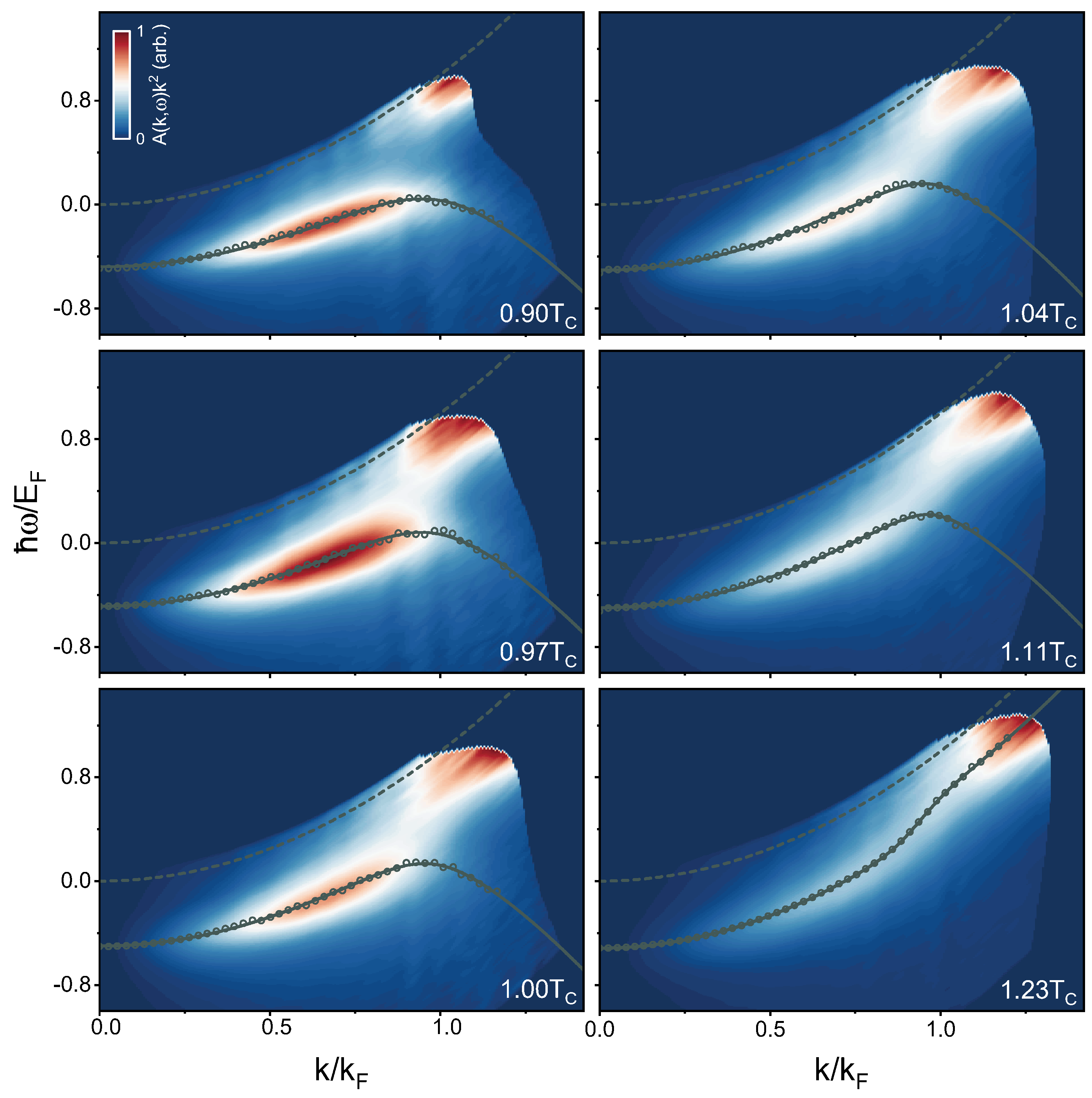}
    \caption{\textbf{Momentum-resolved microwave spectra at various temperatures across the superfluid transition}. 
    In each panel, the spectral intensity is normalized to the maximal intensity. 
    The gray dashed lines denote the free-particle dispersion. 
    The gray circles represent the maximal spectral response $E_{\mathrm{max}}(k)$. 
    For $T \leq 1.11~T_\text{c}$, the two branches are well separated; thus, we show $E_{\mathrm{max}}(k)$ of the lower branch only. 
    The gray solid lines are fitting curves, as described in the text.}
    \label{fig3}
\end{figure*}

    We now investigate how the spectral function $A(\bm{k},\omega)$ varies with temperature and then determine the $T$-dependent pairing gap $\Delta(T)$ across $T_\text{c}$. 
    As shown in Fig.~\ref{fig3}, over a wide temperature range of $0.90T_\text{c} \leq T \leq 1.23T_\text{c}$, the lower and upper branches are clearly distinguishable, with their relative spectral weights varying smoothly across $T_\text{c}$. 
    More importantly, the suppression of spectral weight near the Fermi surface is unequivocally revealed for the normal state at $T\geq T_\text{c}$. 
    This is the spectroscopic smoking-gun signature of a pseudogap in a unitary Fermi gas, without the need to invoke any specific microscopic theories.

    In Fig.~\ref{fig3}, the back-bending of the lower quasiparticle branch can be observed for temperatures up to $1.11T_\text{c}$. 
    We assume that the BCS-like dispersion, Eq.~(\ref{DispersionBCS}), is still applicable above the superfluid transition. 
    The fitted $E_\text{k}^{(-)}$ is shown as the solid lines in Fig.~\ref{fig3}, for $0.90T_\text{c} \leq T \leq 1.11T_\text{c}$. 
    The obtained pairing gap $\Delta(T)$ is plotted as the red triangles in Fig.~\ref{fig4}A, while the quasiparticle effective mass $m^*$ and the Hartree shift $U$ are listed in Table S1\cite{supple}. 
    We find a smooth evolution of $\Delta(T)$ across the superfluid transition, with $\Delta(T_\text{c})=0.280(3)E_\text{F}$ and $\Delta(1.11T_\text{c})=0.195(3)E_\text{F}$. 
    For the spectrum of $T=1.23T_\text{c}$, we can no longer resolve the back-bending in the intensity contour plot, due to the merging of the two branches and the thermally reduced population in the lower branch. 
    The existence of a pseudogap, as clearly revealed by the weak suppression in the spectral weight near the Fermi surface, leads to an ``S-shape'' dispersion, which is more pronounced than that for $1.51T_\text{c}$. 
    By fitting the data to the phenomenological expression of $E_\text{k}$, we obtain $\Delta(1.23T_\text{c})=0.15(4)E_\text{F}$, which is larger than $\Delta(1.51T_\text{c})=0.05(4)E_\text{F}$ as expected. 
    Using linear extrapolation of $\Delta(T)$ down to zero, we may roughly estimate an onset pairing temperature $T^*\gtrsim 1.7T_\text{c}\approx 0.29T_\text{F}$.
    
    \section*{Fermion self energy and energy distribution curve}

\begin{figure*}[h!]
        \centering
        \includegraphics[width =0.9\linewidth]{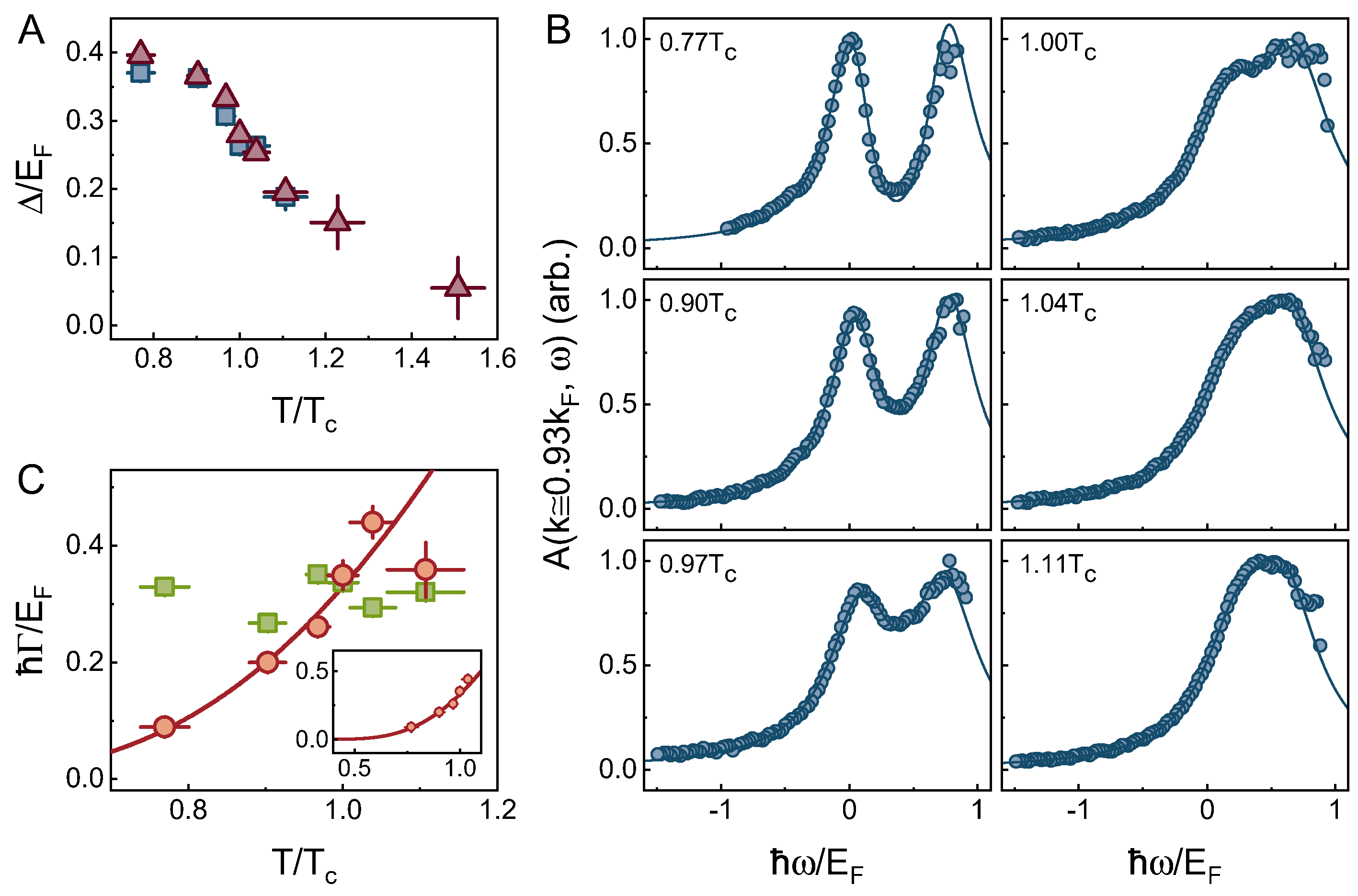}
        \caption{\textbf{Temperature dependence of $\Delta$, $\Gamma$, and the EDCs }.
        (\textbf{A}) The temperature dependence of the pairing gap $\Delta$. Red triangles and blue squares are the results obtained through curve fitting to the locus of $E_\text{max}(k)$ and the EDC, respectively. 
        (\textbf{B}) Normalized EDCs of $A(\bm{k},\omega)$ at $k\simeq k^*\simeq 0.93~k_\text{F}$ (solid circles), accompanied by the fittings to Eq.~(\ref{SelfenergyBCS}) (solid lines) for $0.77T_{\text{c}} \leq T \leq 1.11T_\text{c}$.
        (\textbf{C}) The inverse pair lifetime $\Gamma_0$ (orange circles) and single-particle scattering rate $\Gamma_1$ (green squares) as a function of $T/T_\text{c}$. 
        The solid line corresponds to a thermally activated exponential fitting $\Gamma_0\propto \exp(-E_\text{a}/k_\text{B}T)$ to the data at $T\leq1.04T_\text{c}$. 
        The inset shows the thermally activated behavior over a broad temperature range below $T_\text{c}$. 
        The vertical error bars in (\textbf{A}) and (\textbf{C}) represent one standard deviation, obtained from curve fitting; while the horizontal error bars are the temperature uncertainties.}
        \label{fig4}
    \end{figure*}

    Next, we extract from the spectral function the fermion self energy $\Sigma(\bm{k},\omega)$, a key quantity that characterizes the many-body interaction effects of a unitary Fermi gas (or any interacting system). 
    In Fig.~\ref{fig4}B, we present the normalized EDCs at $k^*\simeq 0.93k_\text{F}$ for $T\leq1.11T_\text{c}$, where the back-bending point $k^*=\sqrt{2m^*(\mu-U)}/\hbar$ is essentially temperature independent. 
    Remarkably, there are two sharp peaks at $\hbar\omega=E^{-}$ and $E^{+}$ in the EDCs for $T<T_\text{c}$, corresponding to the well-defined quasiparticle energies. 
    The dip between two peaks, located at $\hbar\omega\simeq\mu$, marks the maximal suppression of the spectral weight. 
    Thus, in the superfluid phase, the pairing gap $\Delta$ can be easily read off as the half separation of two peaks $(E^+-E^-)/2$. 
    The temperature dependence of EDC is evident as $T$ is raised: (i) the two quasiparticle peaks move toward each other, resulting in a slow decrease in $\Delta$; and (ii) the peak width increases gradually, accompanied by an increasing intra-gap spectral weight. 
    At $T_\text{c}$, the two peaks are significantly broadened, leaving only a weak suppression of the spectral weight near the chemical potential $\hbar\omega \sim \mu$. 
    For higher temperatures $T=1.04T_\text{c}$ and $1.11T_\text{c}$ in the normal state, quasiparticle peaks cannot be identified in the EDCs, and we observe a flat top instead. 
    We attribute the disappearance of the double-peak structure to the significantly increased peak width, which becomes comparable to or larger than the size of the pseudogap at $T>T_\text{c}$.
    
   The measured EDCs can be described using a minimal representation of the fermion self-energy\cite{Norman1998}, which has been widely adopted in the analysis of $A(\bm{k},\omega)$ in high-$T_\text{c}$ superconductors:
    \begin{equation}
            \Sigma(\bm{k},\omega)= \frac{\Delta^2}{\hbar\omega - \mu +\xi(\bm{k})+i\hbar\Gamma_0} - i\hbar\Gamma_1 +  \left[ \xi(\bm{k}) - \epsilon_{\bm{k}} + \mu \right],
            \label{SelfenergyBCS}
    \end{equation}
    where $\xi(\bm{k}) \equiv \hbar^2k^2/2m^*+U-\mu$ with $\xi(\bm{k}^*)=0$, $\Gamma_0$ is responsible for the broadening of quasiparticle peaks due to the finite pair lifetime, and $\Gamma_1$ is the $\omega$-independent single-particle scattering rate that characterizes the contribution from the ``incoherent'' background\cite{Chen2009a}. 
    Using Eq.~(\ref{SelfenergyBCS}), we fit the data with the explicit form of spectral function $A(\bm{k},\omega) = -\operatorname{Im}\Sigma(\bm{k},\omega)/\pi\{[\hbar \omega \\-\epsilon_{\bm{k}} -\operatorname{Re}\Sigma(\bm{k},\omega)]^2+[\operatorname{Im}\Sigma(\bm{k},\omega)]^2\}$. 
    As shown in Fig.~\ref{fig4}B, the fitting curves (solid lines) agree well with the experimental data, enabling us to accurately extract $\Delta$, $\Gamma_0$ and $\Gamma_1$. 
    The obtained pairing gap $\Delta$ is denoted by blue squares in Fig.~\ref{fig4}A. 
    We find $\Delta(1.11T_\text{c})=0.195(3)E_\text{F}$, which agrees well with that extracted from the dispersion fitting (red triangles). 
    Indeed, the two independent approaches, i.e., dispersion fitting with the BCS-like formula and EDC fitting, yield nearly the same pairing gap $\Delta$ at all temperatures. 
    This ensures the quantitative accuracy of the obtained pseudogap size, and thus provides the first experimental benchmark for many-body theories.

    We present $\Gamma_0$ (orange circles) and $\Gamma_1$ (green squares) in Fig.~\ref{fig4}C. 
    The inverse pair lifetime $\Gamma_0$ exhibits a rapid increase with $T$ below $T_\text{c}$, following a thermally activated exponential behavior, $\hbar\Gamma_0 \propto \exp(-E_\text{a}/k_\text{B}T)$, where $E_\text{a}= 2\Delta_0$ with $\Delta_0=0.39(8)E_\text{F}$ being the fitted low-$T$ pairing gap. 
    Note that, the obtained $\Delta_0$ is consistent with our measured pairing gap at 0.77$T_\text{c}$ and a many-body theoretical calculation\cite{Haussmann2007}. 
    This result can be physically understood within a simple picture: $\Gamma_0$ is dominated by a virtual pair breaking and recombination process, which requires an excitation energy of $2\Delta_0$.  
    At $T\gtrsim T_\text{c}$, $\Gamma_0$ approaches approximately $0.4E_\text{F}/\hbar$. 
    The single-particle scattering rate $\Gamma_1$ exhibits a weak temperature dependence across the superfluid transition, with $\hbar\Gamma_1\simeq 0.3E_\text{F}$. 
    This suggests that the single-fermion scattering process is insensitive to the fermion pairing, and thus is largely unchanged in the experimental temperature regime. 
    Indeed, such a weak $T$-dependence of $\Gamma_1$ has also been observed in high-$T_\text{c}$ superconductors. 
    We find $\Gamma_1\simeq 0.3E_\text{F}/\hbar$ at $T_\text{c}$, approximately 1.7 times the Planckian scattering limit\cite{Zaanen2004} (i.e., $k_\text{B}T/\hbar$). 
    This large $\Gamma_1$ is presumably due to the diverging scattering length in the unitary limit. 
    Moreover, the viscous relaxation rate $\tau_{\eta}^{-1}$ of a unitary Fermi gas in the normal state is given by the ratio of pressure $P$ to shear viscosity $\eta$. 
    Near $T_\text{c}$, $P \sim 0.22nE_\text{F}$ and $\eta \sim 1.2n\hbar$, which leads to $\tau_{\eta}^{-1} \sim 0.2E_{\text{F}}/\hbar$, in reasonable agreement with $\Gamma_1 \sim 0.3E_\text{F}/\hbar$, taking $\Gamma_1$ as the transport relaxation rate\cite{Li2022}.

    \section*{Summary}
    
    We establish the existence of a pseudogap, which originates from strong pair fluctuations, in a homogeneous unitary Fermi gas through momentum-resolved microwave spectroscopy in the absence of the final-state effect. 
    We obtain quantitatively accurate $T$-dependent pairing gap $\Delta(T)$, by both analyzing the quasiparticle energy dispersion and fitting the energy distribution curve with a minimal fermion self-energy model. 
    Moreover, the inverse pair lifetime $\Gamma_0$ and the single-particle scattering rate $\Gamma_1$ are successfully extracted, which are two essential quantities for characterizing the microscopic interaction processes in strongly-interacting quantum systems. Our findings not only demonstrate that a many-body pairing pseudogap phase is a precursor to superfluidity, but also provide valuable microscopic details of unitary Fermi gases in the superfluid and normal phases. The quantitative theoretical explanation of the observed pseudogap constitutes a challenge for microscopic quantum many-body theories. Furthermore, our momentum-resolved microwave spectroscopy offers a powerful tool for probing many elusive quantum phases of strongly-interacting fermions, such as pseudogap and $d$-wave superconductivity in the doped repulsive Fermi Hubbard model\cite{Esslinger2010,Bloch2012,Hart2015,Mazurenko2017} and the Fulde-Ferrell-Larkin-Ovchinnikov state\cite{Kinnunen2018} in a spin-polarized Fermi gas.
  
\bibliographystyle{Science}
\bibliography{main}

\textbf{Funding:} This work is supported by the National Key R\&D Program of China (Grant No. 2018YFA0306501), NSFC of China (Grant No. 11874340), the Innovation Program for Quantum Science and Technology (Grant No. 2021ZD0301900), the Chinese Academy of Sciences (CAS), the Anhui Initiative in Quantum Information Technologies, and the Shanghai Municipal Science and Technology Major Project (Grant No.2019SHZDZX01).

\end{document}